\documentclass[11pt]{article}
\usepackage[margin=1in]{geometry}
\usepackage{cite}
\usepackage{subcaption}

\usepackage{standalone}
\usepackage{lineno}
\modulolinenumbers[5]

\usepackage{url}
\usepackage{mathrsfs}

\usepackage{etex}

\usepackage{array}
\usepackage{booktabs}
\usepackage{verbatim}
\usepackage{fancyvrb,relsize}

\usepackage{latexsym}

\newcommand{\fbf}{\textbf}

\usepackage[inline]{enumitem}
\newlist{myenumerate}{enumerate*}{1}
\setlist[myenumerate]{label=\emph{(\arabic*)}, after={.}, itemjoin={{; }}, itemjoin*={{; and }}}

\newcommand{\citep}{\cite}
\newcommand{\shortcitep}{\cite}

\DeclareMathAlphabet{\mathsl}{OT1}{ptm}{m}{sl}

\usepackage{xspace}
\newcommand{\etal}{{et al.\@\xspace}}

\usepackage{colortbl}
\usepackage[table]{xcolor}

\usepackage{tikz}
\usetikzlibrary{arrows,shapes,snakes,automata,backgrounds,petri}
\usetikzlibrary{calc}
\usetikzlibrary{fit}
\usetikzlibrary{matrix}
\usetikzlibrary{positioning}
\usetikzlibrary{decorations}
\usetikzlibrary{decorations.markings}
\tikzstyle{box}=[thin,draw,align=center,anchor=center,minimum height=20,inner sep=2pt,font=\small]
\tikzstyle{infrastructure}=[box,fill=blue!10!gray!30,draw=none,minimum
    height=6ex,rounded corners]
\tikzstyle{middleware}=[infrastructure,fill=blue!20!gray!10]
\tikzstyle{agent}=[box,draw=none,circle,fill=green!10!blue!30!gray!60,minimum width=6]
\tikzstyle{resource}=[box,draw=none,rectangle,fill=green!10!blue!30!gray!60,minimum width=8]
\tikzstyle{scloud}=[shape=cloud,align=center,thick,aspect=4,fill=green!10!blue!30!gray!40]
\tikzstyle{annotation}=[blue!40!black,draw=none,font=\small\sffamily]
\tikzstyle{edge_label}=[annotation,midway,anchor=south,align=center]
\tikzstyle{every text node part/.style}=[align=center]
\tikzstyle{every even column/.style}=[anchor=base east]
\tikzstyle{every odd column/.style}=[anchor=base east]

\usetikzlibrary{chains}

\begin{document}
\pagestyle{plain}

\title{Violable Contracts and Governance for Blockchain Applications}

\author{Munindar P.~Singh \\ Department of Computer Science \\ North Carolina State University \\ \texttt{singh@ncsu.edu} \and  Amit K.~Chopra \\ School of Computing and Communications \\ Lancaster University \\ \texttt{amit.chopra@lancaster.ac.uk}}

\date{}

\maketitle

\begin{abstract}
  We examine blockchain technologies, especially smart contracts, as a
  platform for decentralized applications.  By providing a basis for
  consensus, blockchain promises to upend business models that
  presuppose a central authority.  However, blockchain suffers from
  major shortcomings arising from an over-regimented way of organizing
  computation that limits its prospects.  We propose a sociotechnical,
  yet computational, perspective that avoids those shortcomings.  A
  centerpiece of our vision is the notion of a declarative, violable
  contract in contradistinction to smart contracts.  This new way of
  thinking enables flexible governance, by formalizing organizational
  structures; verification of correctness without obstructing
  autonomy; and a meaningful basis for trust.
\end{abstract}

\fbf{Keywords:} Blockchain; Smart contracts; Contracts; Sociotechnical systems


\section{Introduction}
\label{sec:Introduction}


Blockchains have become prominent in the computing landscape.  The
idea of blockchains originated from the cryptocurrency Bitcoin
\citep{Nakamoto-08:Bitcoin}.  Whereas previous approaches to digital
currency relied upon a central entity to address integrity,
specifically, avoiding double spending, Bitcoin ensures integrity
without a central entity.

Blockchains provide a distributed, shared ledger by bringing together
cryptographic hash functions to achieve immutability and Byzantine
fault tolerance to achieve consensus---as to the definitive current
state of the ledger---among mutually untrusting peers.  Such robust
consensus could enable any digital transaction involving parties who
may not fully trust each other.

By providing a consistent system-wide view of events under weak trust
assumptions, blockchain can enable decentralized applications, for
which lack of trust between participants is a major obstacle.  That's
how blockchain could upend business models in just about any sector:
healthcare, manufacturing, and others
\citep{Yuan+17:blockchain-EHR,Magazzeni+17:smart-contracts},
especially in combination with technologies such as the IoT
\citep{Christidis+Devetsikiotis-16:IoT}.


Consider electronic health records (EHRs) as a representative
application for blockchain.  Healthcare involves multiple stakeholders
with multiple overlapping business relationships, some of a few hours
(emergency room) and some stretching for decades.  Yuan {\etal}
\shortcitep{Yuan+17:blockchain-EHR} motivate the necessity of
integrity, provenance tracking, portability, availability, and access
control.  Among the known data representations, only blockchain
effectively provides these properties.

The theses of this paper are these.  First, decentralized applications
are naturally understood in terms of interactions between autonomous
parties.  Second, blockchain technologies as traditionally construed
are inadequate for decentralized applications.  Third, a perspective
from \emph{sociotechnical systems (STSs)} helps address these
inadequacies via
\begin{myenumerate}
\item an autonomy-preserving representation for \emph{violable}
  contracts
\item an architecture of organizations that balances flexibility and
  rigor to engender trust
\end{myenumerate}

\paragraph{Scope and contributions}
This paper focuses on the challenges relating to decentralized
applications understood as STSs, deemphasizing concerns such as
confidentiality and performance.  Its main contributions are these:
\begin{itemize}
\item An analysis of opportunities for blockchain as a platform for
  decentralized applications and how they are stymied by fundamental
  shortcomings of blockchain.
\item A vision of research challenges necessary to address those
  shortcomings from a sociotechnical perspective.
\end{itemize}

This paper is organized as follows.  Section~\ref{sec:Conceptual}
introduces the relevant conceptual grounding for blockchains and
allied concepts, such as smart contracts.
Section~\ref{sec:Limitations} exposes key limitations of blockchains
for our present purposes.  Section~\ref{sec:MAS-to-BC} describes
research opportunities in addressing those limitations.
Section~\ref{sec:Discussion} concludes with a discussion of prospects.

\section{Blockchain, Conceptually}
\label{sec:Conceptual}

Blockchain solves the longstanding distributed computing problem of
achieving immutable agreement as to the current state of the system,
despite failures and malice---as long as a majority of the computing
power on the network remains in the hands of benevolent participants.
Specifically, blockchain determines the indisputable order in which
events have occurred.


We consider Bitcoin \citep{Nakamoto-08:Bitcoin}, the original
blockchain formulation, to introduce key ideas.  However, we elide
Bitcoin-specific details, such as transaction costs and currency
mining.  Bitcoin is geared toward handling financial transactions, and
provides a ledger of who owns how much of its designated currency,
namely, bitcoins.

Any blockchain consists of a series of blocks, each pointing to, and
including the cryptographic hash of, its predecessor.  The series ends
at the \emph{genesis} block.  Any peer can verify a block's integrity
by verifying its hash.  Since a block contains a pointer to the
previous block, the entire blockchain can be verified.

Each block includes a list of transactions, each transaction being a
record of transfer of ownership of money.  In Bitcoin, the party
providing the coins signs the transaction, indicating consent.  If a
transaction is added to a block, then according to any chain that
extends this block, the specified coins transfer to the recipient.

Whenever a peer receives a transaction, from a client or forwarded
from a peer, it verifies the transaction by checking the signatures
and establishing that the transferrer owns the requisite coins through
previous transfers.  Next, it forwards that transaction to its peers,
who verify and forward it along to their peers, and so on.  Therefore,
potentially, every peer can learn of a new transaction.

Every so often, a peer may build a new block consisting of some or all
of the transactions it has received, by inserting them in a new block.
It creates a block header containing a hash of these transactions and
a hash of the ``previous'' (front of the current \emph{active} chain)
block.

Distribution may lead to competing blocks extending the same
predecessor.  Bitcoin introduced \emph{proof of work} to combat such
branching and promote consensus.  Every block header includes
designated bits called a \emph{nonce}.  Given the rest of the header,
the ``work'' is to produce a nonce such that the hash of the entire
block header satisfies a criterion, namely, being smaller than a
systematically determined number, called the \emph{target}.  The
target adjusts up or down to make block creation or \emph{mining}
easier (faster) or harder (slower), respectively.

Upon mining a block, a peer broadcasts it, in essence proposing to
extend the blockchain.  Any recipient can verify the block's integrity
by checking that its hash is within the target.  If the incoming block
extends the (known) active chain, it extends it and treats it as
active.  Otherwise, it treats the longest of the known branches as
active.  Other features of Bitcoin, including how coins are generated
and miners (peers who create blocks) rewarded, are not relevant here.


The cryptographic hashes in Bitcoin's consensus protocol yield
immutability: any change to the contents of a prior block would
invalidate its hash.  Therefore, an attacker would need to engender
consensus on an alternative active chain.  Assuming the difficulty of
inverting cryptographic hashes, it is not feasible to come up with a
block whose hash equals the hash value stored in its successor block.
And, the consensus mechanism ensures that an attacker cannot fabricate
a new blockchain with different hash values in the chain, as
alternative facts to the ``real'' blockchain---unless the attacker
controls a large fraction of the computational resources in the
network.

\subsection{Smart Contracts}
The notion of a smart contract \citep{Szabo-97:smart} predates
blockchain.  In general, a smart contract specifies contractual
conditions programmatically, such that the contract would
automatically execute when input data meets the stated conditions.
Szabo \citep{Szabo-97:smart} characterizes a vending machine as a
smart contract that takes in coins and outputs a product.  Smart
contracts could potentially be attached to any real-world object
\citep{Christidis+Devetsikiotis-16:IoT}, e.g., a house for rent.

Here, a smart contract is placed on the blockchain, digitally signed
by its creator, with its conditions specified in a program.  Since a
smart contract is public, the parties wishing to exercise it can know
in advance how it will function---provided they understand it.  Hence,
smart contracts can enable commerce in an open setting.

Bitcoin transactions are a simplified form of smart contracts, since
Bitcoin's limited language allows little more than verifying
signatures.  But newer approaches, including Ethereum, ambitiously
support Turing-complete languages for smart contracts that initiate
transactions based on observed events.

\subsection{Permissioned Blockchains}

Bitcoin exposes every transaction in a public ledger.  Moreover,
Bitcoin is slow: it auto-tunes the target so a new block is mined no
more often than once in 10 minutes, the amount of work required being
instrumental in achieving consensus in Bitcoin's untrusted setting.

\emph{Permissioned} blockchains, such as Hyperledger
\citep{Hyperledger-17}, address these shortcomings.  A permissioned
blockchain assumes that it is carried out over a restricted network.
Only approved parties may create transactions and smart contracts, and
validate blocks.  In effect, such an architecture gives up the
openness of blockchains but gains in practicality and an ability to
ensure legality of transactions, which are essential for most serious
purposes.  Such transactions arise commonly---a pharmaceuticals
company wouldn't source its medications from an unknown party, and a
physician is legally required to check credentials of recipients of
patient health data.


\subsection{Upcoming Enhancements}
Blockchain seeks to determine a definitive linear order of events
across the entire system.  Linearizing otherwise unrelated events is
superfluous at best and can cause unacceptable performance loss, which
has led to approaches that relax the linear structure into a directed
acyclic graph \citep{Lewenberg+15:blockchain} and change the consensus
mechanics accordingly.  We focus instead on how to construct
applications that avoid the conceptual limitations of blockchains no
matter what form a blockchain takes.

\section{Sociotechnical Limitations of Smart Contracts}
\label{sec:Limitations}

Let's consider the hazards of smart contracts to motivate the
limitations of blockchain.  The Decentralized Autonomous Organization
(DAO) fiasco \citep{Buterin-17:DAO} is telling.  DAO, a venture
funding entity created as a smart contract on the Ethereum blockchain,
was hacked to the tune of \$50M, by exploiting a flaw in the DAO's
smart contract and the underlying Ethereum virtual machine.  The
specific flaw is merely a symptom that it is impossible to establish
correctness for a program in a Turing-complete language.

Interestingly, this mistake was remedied by causing a fork in the
blockchain.  Specifically, several Ethereum users colluded to extend a
prior block as a way to exclude the undesirable transactions,
discarding legitimate ones as well.  (Now there are at least two
versions of Ethereum, though the details don't concern us here.)  Of
course, a fork was possible only because a large fraction of the
active participants agreed to it.  A minority would not be able to
take such remedies.

The success of the fork undermines the very point that motivated
blockchains, namely, their immutability.  For something like DAO, it
may be appropriate to discard several days of legitimate transactions
to avert a loss of \$50M.  But what would the tradeoffs be in
practice?  Would it be fair to discard an hour's worth of real
commerce at the national scale to save \$50M?  We suspect not.  Or, a
less greedy attacker may succeed by causing only small amounts of harm
at a time, for which detection and reversion are infeasible.

On Hyperledger, because of its permissioned nature, the risk is
presumably better contained.  However, errors in smart contracts are
unavoidable and undesirable outcomes would be difficult to reverse.

We now discuss three major shortcomings in the current conception of
smart contracts.

\subsection{Lack of Control}

The independence of participants with respect to their beliefs and
actions is a crucial aspect of decentralization.  Blockchain supports
independence with regard to private beliefs since consensus applies
only to shared events, which is essential for achieving
interoperation.

However, smart contracts fail independence for actions.  They automate
processing, removing control from the participants.  A smart contract
once launched cannot be overridden.  Indeed, no one can even
contemplate overriding a smart contract because it executes
automatically.

How can we reconcile blockchain with autonomy?

\subsection{Lack of Understanding}

Since the meaning of a smart contract is hidden in a procedure, even
though the procedure is public, it is not apparent if it meets
stakeholder requirements, and how it may be exercised by a
participant.  Since blockchains are immutable, any mistake in
capturing requirements cannot be corrected without violating
immutability.  Therefore, a powerful language for smart contracts
placed on a blockchain poses a huge risk \citep{OHara-17:dumb-idea},
as the DAO incident illustrates.

Instead, we need a language in which we can capture the essential
stakeholder requirements directly.  To improve verification, such a
language would be limited in expressiveness.  To enhance confidence in
capturing valid requirements, it would offer constructs close to the
stakeholders' conception.

How can we develop such a language including an appropriate semantics?

\subsection{Lack of Social Meaning}

Any software application involves contact with the real world.  In
some cases, the real world can be readily abstracted out.  Bitcoin,
being designed for cryptocurrency, is \emph{endogenous}, meaning that
bitcoins exist entirely within the blockchain, which can therefore
ensure their integrity.  Bitcoin is an atypical application since it
excludes considerations other than of transactions involving bitcoins.

More commonly, applications such as healthcare and commerce are
entwined with the real world, social or technical.  For example, in
healthcare, surgical equipment may fail or a patient may deny having
been adequately informed when giving consent.  For physical or
communication failures, the possible resolutions lie in the social
sphere, as traditionally handled through contracts and laws.

The DAO hack demonstrated an integrity violation, indicating a
platform failure.  In a decentralized scenario, any response to an
interoperation failure, including a platform failure, \emph{must} be
social.  Indeed, the response to fork the blockchain was social---it's
just that it was an ad hoc and unverifiable response entirely outside
the computational realm.

How can we enhance blockchain and smart contracts with abstractions to
express and compute with social meaning?

\section{Vision: Compacts, Governance, Verification, Trust}
\label{sec:MAS-to-BC}

The foregoing discussion shows that smart contracts are inadequate for
describing interoperation between autonomous parties: they take over
control of decision making, are opaque, and omit social meaning.  We
now describe our vision that avoids these shortcomings and enables
natural interactions between autonomous parties.


\subsection{Declarative Violable Contracts}

We introduce the term \emph{compact}
(\url{https://www.ldoceonline.com/dictionary/compact}) for our
conception of contracts to avoid confusion with both smart contracts
and traditional contracts.

In contrast to a smart contract, a compact is not a program executed
by the blockchain but a \emph{specification of correct behavior}.  A
compact would be stored on the blockchain.  However, a compact is a
computational artifact: its formal semantics determines which
blockchains satisfy and which violate the compact.  A query processor
based on the semantics would determine the state of each instance of a
compact---whether it is expired, satisfied, or violated.

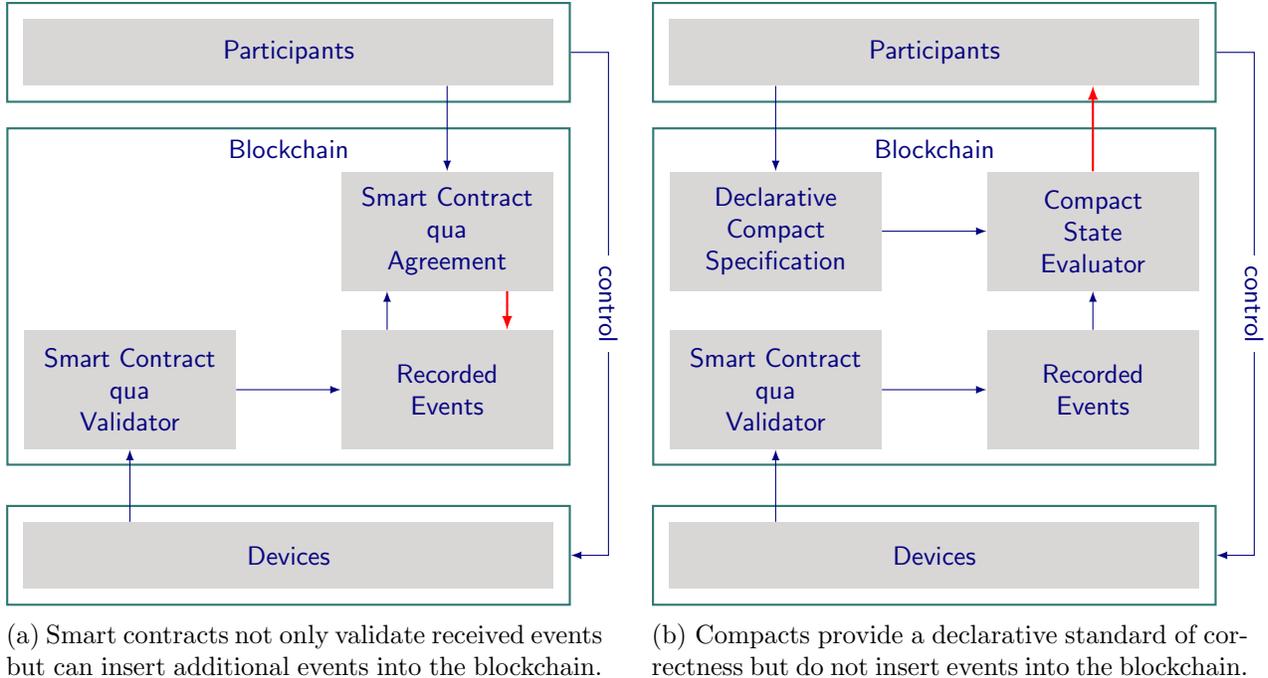
\begin{figure}[htb]
\centering
\begin{subfigure}[t]{0.48\linewidth}
\centering
\definecolor{mpsNavy}{rgb}{0.01,0.01,0.5}
\definecolor{mpsGray}{rgb}{0.85,0.84,0.84}
\definecolor{mpsMyrtleGreen}{rgb}{0.19,0.47,0.45}

\usetikzlibrary{calc}
\usetikzlibrary{fit}
\usetikzlibrary{positioning}
\usetikzlibrary{matrix}

\tikzset{>=latex}		
\begin{tikzpicture}
\tikzstyle{every text node part/.style}=[align=center]

\tikzstyle{module}=[minimum width=80,inner sep=4,font=\small\sffamily,text=mpsNavy,fill=mpsGray,align=center,sharp corners,minimum height=45] 

\tikzstyle{wide}=[minimum width=60,inner sep=0,font=\small\sffamily,text=mpsNavy,fill=mpsGray,align=center,sharp corners,minimum height=25] 

\tikzstyle{emptybox}=[draw=none,fill=none,font=\small\sffamily,text=mpsNavy,minimum height=15]

  \tikzstyle{framing}=[draw=mpsMyrtleGreen,thick,inner sep=6]

\matrix[row sep=0,column sep=40,anchor=center] {

 \node [emptybox] (tl) {}; &
 \node [emptybox] (tr) {}; 
\\[30]

  \node [emptybox] (specs) {};
& \node [module] (eval) {Smart Contract\\qua\\Agreement};
 \\[15]

 \node [module] (validator) {Smart Contract\\qua\\Validator}; &
 \node [module] (events) {Recorded\\Events}; 
\\[0]

 \node [emptybox] (ml) {}; &
 \node [emptybox] (mr) {}; 
\\[10]

 \node [emptybox] (bl) {}; &
 \node [emptybox] (br) {}; 
\\
};

\coordinate (b-spot-x) at ($(specs.center)!0.5!(eval.center)$);
\coordinate (b-spot-y) at ($(eval.north) + (0,0.3)$);

\node [emptybox] at (b-spot-x|-b-spot-y) (blockchain) {Blockchain};

\draw[framing] ($(validator.south west)+(-0.215,-0.2)$) rectangle ($(blockchain.north-|eval.east)+(0.215,0.0)$);

\node [wide,fit=(tl.north-|validator.west)(tr.north-|eval.east)] (p-spot) {};
\node [emptybox] at (p-spot) (participants) {Participants};
\node [framing,fit=(p-spot)(participants)] (p-frame) {};

\node [wide,fit=(bl.south-|validator.west)(br.south-|eval.east)] (d-spot) {};
\node [emptybox] at (d-spot) (devices) {Devices};
\node [framing,fit=(d-spot)(devices)] (d-frame) {};

\draw [mpsNavy,->] (eval.north|-p-spot.south) -- (eval.north);


\draw [mpsNavy,->] (d-spot.north-|validator.south) -- (validator.south);

\draw [mpsNavy,->] (validator) -- (events);

\draw [red,thick,->] (eval.315) -- (eval.315|-events.45);

\draw [mpsNavy,->] (eval.225|-events.north) -- (eval.225);

\draw [mpsNavy,->] (p-frame.east) -- ++(0.5,0) |- (d-frame.east) node[emptybox,fill=white,rotate=-90,pos=0.25] {control};

\end{tikzpicture}
\caption{Smart contracts not only validate received events but can
  insert additional events into the blockchain.}
\label{fig:smart-contracts}
\end{subfigure}
\hfill
\begin{subfigure}[t]{0.48\linewidth}
\centering
\definecolor{mpsNavy}{rgb}{0.01,0.01,0.5}
\definecolor{mpsGray}{rgb}{0.85,0.84,0.84}
\definecolor{mpsMyrtleGreen}{rgb}{0.19,0.47,0.45}

\usetikzlibrary{calc}
\usetikzlibrary{fit}
\usetikzlibrary{positioning}
\usetikzlibrary{matrix}

\tikzset{>=latex}		
\begin{tikzpicture}
\tikzstyle{every text node part/.style}=[align=center]

\tikzstyle{module}=[minimum width=80,inner sep=4,font=\small\sffamily,text=mpsNavy,fill=mpsGray,align=center,sharp corners,minimum height=45] 

\tikzstyle{wide}=[minimum width=60,inner sep=0,font=\small\sffamily,text=mpsNavy,fill=mpsGray,align=center,sharp corners,minimum height=25] 

  \tikzstyle{emptybox}=[draw=none,fill=none,font=\small\sffamily,text=mpsNavy,minimum height=15]

  \tikzstyle{framing}=[draw=mpsMyrtleGreen,thick,inner sep=6]

\matrix[row sep=0,column sep=40,anchor=center] {

 \node [emptybox] (tl) {}; &
 \node [emptybox] (tr) {}; 
\\[30]

  \node [module] (specs) {Declarative\\Compact\\Specification};
& \node [module] (eval) {Compact\\State\\Evaluator};
 \\[15]

 \node [module] (validator) {Smart Contract\\qua\\ Validator}; &
 \node [module] (events) {Recorded\\Events}; 
\\[0]

 \node [emptybox] (ml) {}; &
 \node [emptybox] (mr) {}; 
\\[10]

 \node [emptybox] (bl) {}; &
 \node [emptybox] (br) {}; 
\\
};

\node [emptybox] at ($(specs.north)!0.5!(eval.north) + (0,0.3)$) (blockchain) {Blockchain};

\draw[framing] ($(validator.south west)+(-0.215,-0.2)$) rectangle ($(blockchain.north-|eval.east)+(0.215,0.0)$);

\node [wide,fit=(tl.north-|specs.west)(tr.north-|eval.east)] (p-spot) {};
\node [emptybox] at (p-spot) (participants) {Participants};
\node [framing,fit=(p-spot)(participants)] (p-frame) {};

\node [wide,fit=(bl.south-|specs.west)(br.south-|eval.east)] (d-spot) {};
\node [emptybox] at (d-spot) (devices) {Devices};
\node [framing,fit=(d-spot)(devices)] (d-frame) {};

\draw [mpsNavy,->] (specs.north|-p-spot.south) -- (specs.north);

\draw [red,thick,->] (eval.north) -- (eval.north|-p-spot.south);

\draw [mpsNavy,->] (specs) -- (eval);

\draw [mpsNavy,->] (d-spot.north-|validator.south) -- (validator.south);

\draw [mpsNavy,->] (validator) -- (events);

\draw [mpsNavy,->] (eval.south|-events.north) -- (eval.south);

\draw [mpsNavy,->] (p-frame.east) -- ++(0.5,0) |- (d-frame.east) node[emptybox,fill=white,rotate=-90,pos=0.25] {control};

\end{tikzpicture}
\caption{Compacts provide a declarative standard of correctness but do
  not insert events into the blockchain.}
\label{fig:compacts}
\end{subfigure}
\caption{Comparing compacts and smart contracts in a blockchain architecture.}
\label{fig:architecture}
\end{figure}

Figure~\ref{fig:architecture} illustrates how compacts differ from
smart contracts.  In both settings, participants (social entities) own
and control devices (technical entities), such as computers, sensors,
and vehicles.  The blockchain records events produced by the devices,
upon validation by a smart contract.  In the traditional conception of
Figure~\ref{fig:smart-contracts}, the participants additionally
specify their business agreements as smart contracts that carry out
actions and record events on the blockchain.  In our conception, in
Figure~\ref{fig:compacts}, the participants specify the compacts
corresponding to their business relationships.  Given the recorded
events, the evaluator determines whether a compact is satisfied,
violated, or neither.  It informs the participants about states of
relevant compacts, but does \emph{not} record events.  That is, of the
two functions of smart contracts in Figure~\ref{fig:smart-contracts},
Figure~\ref{fig:compacts} retains only one.

A compact helps balance autonomy and correctness.  A party to a
compact, exercising its autonomy, may violate it.  For example, a
compact in healthcare may specify that a hospital prohibits a nurse to
share a patient's data without the patient's consent.  Yet, a nurse
Bob may share patient Charlie's data with cardiologist Alice without
Charlie's consent.  From the semantics, given recorded events, we can
compute whether the compact was satisfied or violated.  But, of
course, the violation in the present example doesn't entail that Bob
was malfeasant.  It could be that Charlie had a medical emergency and
was in no condition to give consent.  Bob could be rewarded for saving
Charlie's life for his workaround \cite{Koppel+15:workarounds}.
The compacts view highlights the importance of detecting and resolving
conflicting requirements \cite{dos-Santos+17-conflicts}.

To recover understanding, control, and make the social meaning
explicit, we need a declarative representation for compacts that
captures the essence of traditional contracts.  A compact would
explicitly state what each concerned party may expect from the others.
To this end, the formal notion of \emph{norms} yields promising
constructs.  As motivated by Georg von Wright, the father of deontic
logic, this kind of norm expresses regulatory force
\cite{Von-Wright-99:Personal}.  This form of norm is directed from its
subject to its object, and states logical conditions under which it
goes in force and under which it completes \cite{TIST-13-Governance}.
For example, a prohibition is a kind of norm in this sense.

\subsection{Organizations and Governance}

Consensus on what has transpired can support decentralized
applications by averting disputes as to the public facts.  But, as
envisioned here, the parties may nevertheless violate applicable
compacts.

An \emph{organizational context} for a norm is the organization in
which the norm arises \citep{Ailaw-99}.  The context is a principal on
par with any other, and may feature as a subject or object of another
norm.  This simple representation can be valuable: The context can
serve as an adjudicating authority for disputes; norms involving it
help mitigate violations of norms in a compact \citep{Computer-09}.

Let us extend the above patient information example to make the
hospital the context of the prohibition norm and to introduce a
commitment from the hospital to the patient to investigate any sharing
of the patient's data without the patient's consent.  Now when Bob
reveals Charlie's data without Charlie's consent, the hospital's
commitment to Charlie is activated.  The hospital can satisfy the
commitment by conducting its investigation, upon which it may
exonerate and reward Bob or penalize him.

The above example illustrates \emph{governance}, i.e., how coherence
is achieved in interactions, in the absence of a central authority
\citep{Pitt+Artikis-15:open,Frantz+13:nADICO,TIST-13-Governance}.
Governance is a prerequisite for accountability and trust, which are
means with which to balance autonomy and correctness.

Decentralized applications cannot avoid governance: the choice is
whether to leave it ad hoc and manual or to make it formal and
computational, as we envision.  In our conception, every decentralized
application is associated with an organization, which serves as the
context of the compact that defines the application.  Today, we see
nebulous communities, such as the Ethereum network, or somewhat more
crisp organizations such as on permissioned blockchains.  However,
these organizations lack a computational representation such as a
compact.  Consequently, there is no precise characterization of what
an organization can expect from its members and vice versa.

How can we represent and compute with formal organizations in relation
to compacts on blockchain?  How can we treat an organization as an
entity on par with other parties in a decentralized application?

\subsection{Programming and Verifying Interactions}

Achieving coordination is nontrivial in decentralization applications.
Existing approaches hardcode coordination in the participants, i.e.,
their agents.  Doing so reduces flexibility in interoperation and
hides essential details, thereby preventing composing compacts.
Therefore, it is important to specify the coordination declaratively.
Doing so requires not just formal semantics for data
\cite{Third+Domingue-17:ledgers} but also models of causality and
integrity constraints on interactions underlying the data
\citep{AAMAS-BSPL-12}.  To facilitate integrity preservation of
decentralized applications, we can capture information-level integrity
constraints in a smart contract that validates received transactions.
Doing so would prevent entering an information state wherein the
applicable compact's state would be confused.

To enable interoperation, we must formalize how an interaction
proceeds, not just who participates or what data they exchange.
Blockchains, e.g., Hyperledger \shortcitep{Hyperledger-17}, provide
coordination abstractions such as a \emph{channel}---a subnet on which
only participants can access information.  A channel supports
confidentiality and helps decouple participants by hiding irrelevant
information.  In essence, we would enhance channels into formal
\emph{protocols}.  Moreover, to support declarative compacts, we must
specify a protocol declaratively so that the state of a compact can be
computed and satisfaction or violation determined based exclusively on
information in the blockchain  \cite{Shams+17:norms}.

To capture the intuition that a decentralized application is specified
via a compact, we would need to generate protocols automatically from
a compact such that each involves only the relevant participants.
What causality and integrity constraints on protocols arise naturally
from a compact?  What causality and integrity constraints arise from
application-specific considerations such as which parties controls
what events?  And, which pieces of information are generated
atomically, and so on?  Recent work on generating protocols that
guarantee alignment of commitments provides a start
\cite{IJCAI-17:Tosca}.

How can we ensure the protocols yield the information transfer needed
by each party to enact its part of the application and to verify
\emph{compliance} of other parties with the compact?  Notice that a
closed system approach for compliance does not apply in blockchain.
Specifically, contrary to recent suggestions
\citep{Magazzeni+17:smart-contracts}, we \emph{cannot} determine
compliance based on internal details of a participant, such as its
beliefs, intentions, or sincerity.  In contrast, verifiability in an
open setting demands a \emph{public semantics} \citep{Computer-98},
which is what a shared ledger offers---a major point in favor of
blockchain.

\subsection{Meaningful Trust and Reputation}

The autonomy of participants and embedding in the real world suggest
that participants would need to trust one another to interoperate.
The possibility of violation of a compact creates a
\emph{vulnerability}, a hallmark of trust
\cite{Castelfranchi+Falcone-10}.  Blockchain obviates the need for
trust only to the extent that the governance structures provide
assurance against malfeasance by another participant and the
structures themselves are trusted.

However, blockchain can serve as a platform for promoting meaningful
trust between parties.  First, quite naturally, the states of relevant
compacts provide an opportunity to make evidential trust judgments.
Violation and satisfaction of a norm would mean a lowering and
raising, respectively, of trust in the concerned party with respect to
similar norms.  Second, explicit governance engenders trust because
parties that may otherwise not transact with each other would do so
because compacts for governance would give them assurance that
malefactors would be sanctioned.  A party may violate a compact by
failing to satisfy its conditions, but if it does so its violation
would be determinable from the blockchain.  The aggrieved party
\citep{Law-17:aggrieved} may file a complaint, also recorded in the
blockchain, presumably triggering a governance compact.

Third, governance can provide a basis for capturing the trust
assumptions by formalizing what counts as evidence for what norm.
Consensus on blockchain concerns the events observed.  But armed with
a governance structure, we can encapsulate norm-relevant evidence
within an event in a manner that reflects the application semantics.
For example, a norm may rely upon a patient having a benign tumor.
But, in medical practice \cite{ASPE-10}, whether a tumor is benign is
a fact that is established by the tumor board of the hospital.  That
is, the tumor board's assertion counts as the tumor being benign.

\section{Discussion}
\label{sec:Discussion}

The emergence of blockchain as a platform for decentralized
applications exposes new usage scenarios.  Concomitantly arise new
expectations from computing---specifically, in terms of governance
(organizations, norms, privacy) and trust.  Consequently, it becomes
essential to bring forth sociotechnical considerations into computing.

Table~\ref{tab:contrast} highlights how our vision of compacts
contrasts with existing approaches.  Here, the blockchain
declaratively represents contractual relationships; maintains relevant
events; enables a participant to violate a contract if it so desires;
computes whether the contract is satisfied, violated, expired, or
otherwise pending; thereby activating applicable governance contracts
and providing a basis for trust.

\begin{table}[htb!]
\centering
\caption{Contrasting compacts with traditional and smart contracts.}
\label{tab:contrast}
\rowcolors{1}{gray!25}{white}
\begin{tabular}{>{\columncolor[gray]{0.8}}l l l l}
  & \fbf{Traditional} & \fbf{Smart} & \fbf{Compacts} \\
\fbf{Specification} & Text & Procedure & Formal, declarative \\
\fbf{Automation} & None & Full & Compliance checking \\
\fbf{Participants' Control} & Complete & None & Complete \\
\fbf{Venue} & External & Within blockchain & Recorded on blockchain\\
\fbf{Trust Model} & Hidden & Hardcoded & Explicit\\
\fbf{Social Meaning} & Informal & None &  Formal \\
\fbf{Standard of Correctness} & Informal legal & Whatever executes & Formal legal \\
\fbf{Scope} & Open but ad hoc & Closed & Sociotechnical \\
\end{tabular}
\end{table}

In this manner, we envision computational representation and reasoning
about sociotechnical considerations.  Specifically, we advocate
developing approaches for programming interactions on or through the
blockchain that build on and support effective governance and trust.

This vision yields valuable research opportunities concerning how
participants
\begin{myenumerate}
\item preserve autonomy in being able to violate a contract and verify
  each other's compliance
\item deal with events in the real business or social worlds, external
  to the blockchain
\item maximize flexibility in having their interactions minimally
  constrained to interoperate successfully
\item most importantly, build and realize governance structures to
  deal with autonomy and exceptions
\end{myenumerate}


\textbf{Acknowledgments.}
Singh was supported by an IBM Faculty Award and Chopra by EPSRC
grant EP/N027965/1 (\emph{Turtles}).




\bibliographystyle{plain}

\begin{thebibliography}{10}

\bibitem{ASPE-10}
ASPE.
\newblock The importance of radiology and pathology communication in the
  diagnosis and staging of cancer: Mammography as a case study, November 2010.
\newblock Office of the Assistant Secretary for Planning and Evaluation, U.S.\
  Department of Health and Human Services; available at
  \protect\url{http://aspe.hhs.gov/sp/reports/2010/PathRad/index.shtml}.

\bibitem{Law-17:aggrieved}
{Black}.
\newblock Black's law dictionary: Aggrieved party, 2017.
\newblock \url{http://thelawdictionary.org/aggrieved-party/}.

\bibitem{Buterin-17:DAO}
Vitalik Buterin.
\newblock Critical update re: {DAO} vulnerability.
\newblock
  \protect\url{https://blog.ethereum.org/2016/06/17/critical-update-re-dao-vulnerability/},
  June 2016.

\bibitem{Castelfranchi+Falcone-10}
Cristiano Castelfranchi and Rino Falcone.
\newblock {\em Trust Theory: A Socio-Cognitive and Computational Model}.
\newblock Agent Technology. John Wiley \& Sons, Chichester, United Kingdom,
  2010.

\bibitem{Christidis+Devetsikiotis-16:IoT}
Konstantinos Christidis and Michael Devetsikiotis.
\newblock Blockchains and smart contracts for the {Internet} of {Things}.
\newblock {\em IEEE Access}, 4:2292--2303, June 2016.

\bibitem{dos-Santos+17-conflicts}
J{\'{e}}ssica~Soares dos Santos, Jean de~Oliveira~Zahn, Eduardo~Augusto
  Silvestre, Viviane~Torres da~Silva, and Wamberto~Weber Vasconcelos.
\newblock Detection and resolution of normative conflicts in multi-agent
  systems: A literature survey.
\newblock {\em Journal of Autonomous Agents and Multi-Agent Systems (JAAMAS)},
  31(6):1236--1282, November 2017.

\bibitem{Frantz+13:nADICO}
Christopher Frantz, Martin~K. Purvis, Mariusz Nowostawski, and Bastin Tony~Roy
  Savarimuthu.
\newblock {nADICO}: A nested grammar of institutions.
\newblock In {\em Proceedings of the 16th International Conference on
  Principles and Practice of Multi-Agent Systems (PRIMA)}, volume 8291 of {\em
  Lecture Notes in Computer Science}, pages 429--436, Dunedin, New Zealand,
  December 2013. Springer.

\bibitem{Hyperledger-17}
Hyperledger.
\newblock Hyperledger, 2017.
\newblock Linux Foundation: \url{http://hyperledger.org}.

\bibitem{IJCAI-17:Tosca}
Thomas~Christopher King, Ak{\i}n G{\"u}nay, Amit~K. Chopra, and Munindar~P.
  Singh.
\newblock Tosca: Operationalizing commitments over information protocols.
\newblock In {\em Proceedings of the 26th International Joint Conference on
  Artificial Intelligence (IJCAI)}, pages 256--264, Melbourne, August 2017.
  IJCAI.

\bibitem{Koppel+15:workarounds}
Ross Koppel, Sean Smith, Jim Blythe, and Vijay Kothari.
\newblock Workarounds to computer access in healthcare organizations: You want
  my password or a dead patient?
\newblock In Karen~L. Courtney, Alex Kuo, and Omid Shabestari, editors, {\em
  Driving Quality in Informatics: Fulfilling the Promise}, volume 208 of {\em
  Series on Technology and Informatics}, pages 215--220. IOS Press, Amsterdam,
  2015.

\bibitem{Lewenberg+15:blockchain}
Yoad Lewenberg, Yonatan Sompolinsky, and Aviv Zohar.
\newblock Inclusive block chain protocols.
\newblock In {\em Proceedings of the 19th International Conference on Financial
  Cryptography and Data Security (FC)}, volume 8975 of {\em Lecture Notes in
  Computer Science}, pages 528--547, San Juan, Puerto Rico, January 2015.
  Springer.

\bibitem{Magazzeni+17:smart-contracts}
Daniele Magazzeni, Peter McBurney, and William Nash.
\newblock Validation and verification of smart contracts: A research agenda.
\newblock {\em IEEE Computer}, 50(9):50--57, September 2017.

\bibitem{Nakamoto-08:Bitcoin}
Satoshi Nakamoto.
\newblock Bitcoin: A peer-to-peer electronic cash system.
\newblock \url{https://bitcoin.org/bitcoin.pdf}, 2008.

\bibitem{OHara-17:dumb-idea}
Kieron O'Hara.
\newblock Smart contracts -- dumb idea.
\newblock {\em IEEE Internet Computing}, 21(2):97--101, March 2017.

\bibitem{Pitt+Artikis-15:open}
Jeremy~V. Pitt and Alexander Artikis.
\newblock The open agent society: Retrospective and prospective views.
\newblock {\em Artificial Intelligence and Law}, 23(3):241--270, September
  2015.

\bibitem{Shams+17:norms}
Zohreh Shams, Marina~De Vos, Julian Padget, and Wamberto~Weber Vasconcelos.
\newblock Practical reasoning with norms for autonomous software agents.
\newblock {\em Engineering Applications of {AI}}, 65:388--399, October 2017.

\bibitem{Computer-98}
Munindar~P. Singh.
\newblock Agent communication languages: Rethinking the principles.
\newblock {\em IEEE Computer}, 31(12):40--47, December 1998.

\bibitem{Ailaw-99}
Munindar~P. Singh.
\newblock An ontology for commitments in multiagent systems: Toward a
  unification of normative concepts.
\newblock {\em Artificial Intelligence and Law}, 7(1):97--113, March 1999.

\bibitem{AAMAS-BSPL-12}
Munindar~P. Singh.
\newblock Semantics and verification of information-based protocols.
\newblock In {\em Proceedings of the 11th International Conference on
  Autonomous Agents and MultiAgent Systems (AAMAS)}, pages 1149--1156,
  Valencia, Spain, June 2012. IFAAMAS.

\bibitem{TIST-13-Governance}
Munindar~P. Singh.
\newblock Norms as a basis for governing sociotechnical systems.
\newblock {\em ACM Transactions on Intelligent Systems and Technology (TIST)},
  5(1):21:1--21:23, December 2013.

\bibitem{Computer-09}
Munindar~P. Singh, Amit~K. Chopra, and Nirmit Desai.
\newblock Commitment-based service-oriented architecture.
\newblock {\em IEEE Computer}, 42(11):72--79, November 2009.

\bibitem{Szabo-97:smart}
Nick Szabo.
\newblock Formalizing and securing relationships on public networks.
\newblock {\em First Monday}, 2(9), September 1997.

\bibitem{Third+Domingue-17:ledgers}
Allan Third and John Domingue.
\newblock Linked data indexing of distributed ledgers.
\newblock In {\em Proceedings of the 26th International Conference on World
  Wide Web Companion}, pages 1431--1436, Perth, 2017. ACM.

\bibitem{Von-Wright-99:Personal}
Georg~Henrik Von~Wright.
\newblock Deontic logic: A personal view.
\newblock {\em Ratio Juris}, 12(1):26--38, March 1999.

\bibitem{Yuan+17:blockchain-EHR}
Ben Yuan, Wendy Lin, and Colin McDonnell.
\newblock Blockchains and electronic health records.
\newblock Unpublished note: \url{http://mcdonnell.mit.edu/blockchain_ehr.pdf},
  2017.

\end{thebibliography}

\end{document}